\documentclass[preprint,showpacs,preprintnumbers,amsmath,amssymb, superscriptaddress,longbibliography,nofootinbib]{revtex4-1}
\usepackage{graphicx} % Required for inserting images
\usepackage{xcolor}
\usepackage{amsmath}
 \usepackage{hyperref}
\usepackage[top = 2cm, bottom = 2cm, right = 2cm, left =2cm]{geometry}

\begin{document}

\title{Braneworld Black Bounce to Transversable Wormhole}

\author{Tiago M. Crispim \footnote{E-mail: tiago.crispim@fisica.ufc.br}}\affiliation{Departamento de F\'isica, Universidade Federal do Cear\'a, Caixa Postal 6030, Campus do Pici, 60455-760 Fortaleza, Cear\'a, Brazil}

\author{Milko Estrada \footnote{E-mail: milko.estrada@gmail.com}}\affiliation{Facultad de Ingeniería y Empresa, Universidad Católica Silva Henríquez, Chile}

\author{C. R. Muniz\footnote{E-mail: celio.muniz@uece.br}}\affiliation{Universidade Estadual do Cear\'a (UECE), Faculdade de Educa\c c\~ao, Ci\^encias e Letras de Iguatu, 63500-000, Iguatu, CE, Brazil.}

\author{G. Alencar \footnote{E-mail: geova@fisica.ufc.br}}\affiliation{Departamento de F\'isica, Universidade Federal do Cear\'a, Caixa Postal 6030, Campus do Pici, 60455-760 Fortaleza, Cear\'a, Brazil}

\date{\today}

\begin{abstract}
We provide a way for embedding a 4-dimensional geometry corresponding to the Simpson-Visser (SV) spacetime—which is capable of representing a traversable wormhole, a one-way wormhole, or a regular black hole—into a Randall-Sundrum setup. To achieve this, we linearly deform the bulk geometry and the bulk matter distribution concerning a coupling constant. These deformations induce a transition from a $5D$ vacuum AdS state to an anisotropic matter distribution. The latter results in the induced geometry on the brane transitioning from a singular Schwarzschild spacetime to a regularized SV spacetime. Since there are no sources or matter fields on the brane, we can assert that the induced SV geometry on the brane arises from the influence of geometrical and matter deformations in the bulk. Thus, the central singularity is suppressed. We determine the cases where the energy conditions are either satisfied or violated. Our spacetime is asymptotically radial AdS, which is intriguing given the absence of a global AdS box that would prevent instability under larger wavelength perturbations. Therefore, it is no longer appropriate to claim that instability exists for very small perturbations near the AdS horizon. Thus, we propose that the stability of the solution can be analyzed by examining the speed of sound due to the presence of matter fields in the energy momentum tensor.
\end{abstract}

\maketitle

\section{Introduction}

Over the last few years, General Relativity(GR) has gained significant attention. This is due to the present era of high-precision measurements of black holes and cosmology, which allowed the direct measurement of gravitational waves by LIGO and VIRGO, the image of supermassive black holes by EHT, and precise measurements of the $\Lambda CDM$ parameters\cite{LIGOScientific:2016aoc, EventHorizonTelescope:2022wkp, EventHorizonTelescope:2019dse, LIGOScientific:2017vwq,Planck:2018vyg,ACT:2020gnv,SDSS:2006srq}. This opens a new window to detect extra dimensions, and the Randall-Sundrum model (RS) have garnered increased interest. It has many attractive features, explaining why gravity is weaker than the other forces of nature, the hierarchy problem, beyond being an alternative to compactification\cite{Randall:1999ee,Randall:1999vf}. RS models have also garnered interest in testing the potential existence of extra dimensions using data obtained from gravitational experiments \cite{Visinelli:2017bny,Vagnozzi:2019apd}.

The RS model involves the embedding of $4D$ flat branes, {\it i.e.}, whose geometry corresponds to Minkowski spacetime. Since its inception, various generalizations of the Randall-Sundrum model, featuring different types of four-dimensional geometries including black holes and wormholes, have emerged. The approach to studying these generalizations is primarily based on two main approximations:  one originates from the brane, and the other from the bulk. In the former, the four-dimensional black holes or wormholes are constructed by studying the equations of motion induced on the brane \cite{Shiromizu:1999wj}. Some examples in references \cite{Molina:2012ay,Neves:2015vga,Neves:2012it,Casadio:2015gea}. However, this methodology does not fully allow for testing the influence of bulk geometry on the physical and geometric behavior of the four-dimensional geometry. Regarding the latter, the second approach constructs spacetimes directly from the bulk, and the brane equations are derived from there. See \cite{Chamblin:1999by,Kanti:2002fx,Creek:2006je}. The approach embraced in this study aligns with the latter. See other recent and different applications to braneworld models in references \cite{Nakas_2020,Nakas:2023yhj,Neves:2021dqx}.

It is important to mention that analysis of observations of the shadows of M87* has provided indications that the induced geometry on the brane could represent both ultra-compact objects and wormholes. In particular, reference \cite{Banerjee:2022jog}, studying the embedding of such objects, demonstrated that the braneworld scenario fits better with the observed shadow and the diameter of the image of Sgr A* compared to the corresponding situation in general relativity. Even though wormholes, in general, require exotic matter to remain stable, in the presence of extra dimensions such is not the case. Therefore, they showed that the wormhole solution considered does not require exotic matter {\it on the brane}. Thus, this reason renewed the interest in embedding other types of geometries different from the Schwarzschild vacuum solution, such as wormholes or regular black holes.

 In another direction, a theoretical value of the AdS radius was estimated in the RS framework from recent data from near-simultaneous detection of the GW event GW170817 from the LIGO/Virgo collaboration and its electromagnetic counterpart, the short gamma-ray burst GRB170817A detected by the Fermi Gamma-ray Burst Monitor and the International Gamma-Ray Astrophysics Laboratory Anti-Coincidence Shield spectrometer, \cite{Visinelli:2017bny}. This estimation revealed that the results obtained are inconsistent with observations of the solar system at millimeter scales. Therefore, the bulk could be described by a source other than a vacuum represented by an AdS cosmological constant. 

On the other hand, in reference \cite{Simpson_2019}, a method for regularizing the Schwarzschild spacetime was proposed by introducing a regularization parameter $a^2$. Depending on the value of $a$, the four-dimensional spacetime could represent a traversable wormhole (TWH), a one-way wormhole (OWWH) with an extremal null throat at $r=0$, or a regular black hole (RBH). The method has been used recently to regularize cylindrical black holes, such as BTZ and $4d$ black strings\cite{Furtado:2022tnb, Lima:2022pvc, Lima:2023arg, Lima:2023jtl, Alencar:2024yvh, Bronnikov:2023aya}.  Therefore, it is interesting to investigate whether the same is true for $5D$ black strings. Another feature of the SV solution is that a source is necessary; therefore, it is not a vacuum solution. This is very attractive since, as said above, the bulk must be described by a source other than a vacuum. 

Finally, in recent years, there has been considerable attention drawn to the fact that the deformation of the Einstein tensor and the matter content within the energy-momentum tensor can modify the geometry of an object \cite{Ovalle:2017fgl}. 
In this work, we will study the geometric characteristics that the five-dimensional bulk and the energy-momentum distribution in the five-dimensional bulk must possess for the induced geometry on the brane to correspond to TWH, an OWWH, or an RBH in the SV set-up. We analyze the locations where the solution is regular and investigate the energy conditions. We discuss the effects of deformations in the geometry and matter of the $5D$ bulk on the induced $4D$ geometry on the brane and speculate about the nature of the new sources of matter derived from this deformation. The stability of the obtained solutions will be studied considering the sound speed within the source on the brane, as well as the potential presence of an AdS box that may prevent instability under wavelength perturbations.

\section{Simpson-Visser regularization}

In order to represent the embedding of a singularity-free object in a braneworld setup, we can consider that this object corresponds to what results from applying the Simpson-Visser regularization \cite{Simpson_2019} to a $4D$ Schwarzschild space. Thus, we will write the $5D$ line element as:
\begin{equation} \label{ecuacionInicial}
     ds^2 =\frac{l^2}{z^2}\left[- \mathcal{U}(r)dt^2 + \mathcal{U}(r)^{-1}dr^2 + (r^2 + a^2)(d\theta^2 + \sin^2\theta d\phi^2) + dz^2\right],
\end{equation}
where $\mathcal{U}(r) = 1 - 2M/\sqrt{r^2 + a^2}$ and the coordinates now take on values in the ranges $t \in (-\infty, +\infty)$, $r \in  [0, +\infty)$, $\phi \in [0,2\pi]$, $\theta \in [0, \pi]$ and $z \in (-\infty, +\infty)$.

We can analyze the causal structure of this geometry by taking $ \theta = \pi/2$, $\phi = 0$, $z = cte = z_0$, and $ds^2=0$, from which we obtain the following expression for $dr/dt$
\begin{equation}
    \frac{dr}{dt} = \pm \left(1 - \frac{2M}{\sqrt{r^2 + a^2}}\right).
\end{equation}
This is the same expression obtained by \cite{Simpson_2019}. Thus, the conclusions are similar, i.e.:
\begin{itemize}
    \item If $a > 2M$, then for any value of $r$, $\frac{dr}{dt} \neq 0$, and therefore we have a braneworld two-way traversable wormhole;
    \item If $a = 2M$, $\frac{dr}{dt} \to 0$ when $r \to 0$, which shows that at $r = 0$ we have a horizon. Thus, we have a braneworld one-way wormhole.
    \item If $a < 2M$, $\frac{dr}{dt} = 0$ for $r_h = \pm \sqrt{(2M)^2 - a^2}$, and so we have a braneworld regularized black hole with the horizon at $r_h$. 
\end{itemize}
As a simplification, we can also make a coordinate change in the radial direction given by $\tilde{r}^2 = r^2 + a^2$. With this, the metric takes the form
\begin{equation} \label{elementodelineafinal1}
     ds^2 =\frac{l^2}{z^2}\left[-\left(1 - \frac{2M}{\tilde{r}}\right)dt^2 + \frac{d\tilde{r}^2}{\left(1 - \frac{2M}{\tilde{r}}\right)\left(1 - \frac{a^2}{\tilde{r}^2}\right)} + \tilde{r}^2(d\theta^2 + \sin^2\theta d\phi^2) + dz^2\right],
\end{equation}
where now $t \in (-\infty, +\infty)$, $\tilde{r} \in  [a, +\infty)$, $\phi \in [0,2\pi]$, $\theta \in [0, \pi]$ and $z \in (-\infty, +\infty)$.

In another direction, for the non-zero components of the Einstein tensor, we have
\begin{eqnarray}
   G^0_0 &=& \frac{1}{l^2}\left(6 + \frac{a^2z^2(\tilde{r} - 4M)}{\tilde{r}^5}\right); \;\; G^1_1 = \frac{1}{l^2}\left(6 - \frac{a^2z^2}{\tilde{r}^4}\right); \nonumber\\
   G^2_2 &=& G^3_3 = \frac{1}{l^2}\left(6 +\frac{a^2z^2(\tilde{r} - M)}{\tilde{r}^5}\right); \;\;\;G^4_4 = \frac{1}{l^2}\left(6 + \frac{a^2z^2(\tilde{r} - 3M)}{\tilde{r}^5}\right). \label{ET}
\end{eqnarray}
Note that all components of the Einstein tensor can be written as:
\begin{equation} \label{EinsteinDeformado}
    G^A_B = (G^A_B )_{\text{seed}} + a^2 \cdot (\text{Geometric Deformation}) \delta^A_B.
\end{equation}
 Therefore, the coupling $a^2$ induces a linear deformation of the bulk geometry, with the $5D$ Hawking black string \cite{Chamblin:1999by} as a seed solution. This also shows that a linear deformation of the $5D$ energy-momentum tensor is necessary (as detailed below in equation \ref{pressures}).
 \begin{equation}\label{MateriaDeformado}
      T^A_B = (T^A_B )_{\text{seed}} + a^2 \cdot (\text{Matter Deformation}) \delta^A_B.
 \end{equation}
 This results from the induced geometry on the brane transitioning from a Schwarzschild singular spacetime to a regularized SV spacetime. It can be interpreted as a matter deformation, generating a transition from a vacuum distribution, characterized by the $5D$ cosmological constant, to an anisotropic distribution. It is worth mentioning that as the radial coordinate tends to infinity, all components of the geometrical deformation of the Einstein tensor vanish. This is consistent with the observation that, in this limit, spacetime behaves as AdS$_5$. 
 
 It is worth noting that linear deformations of geometry and matter have attracted considerable attention in recent years through the Gravitational decoupling algorithm \cite{Ovalle:2017fgl}, where, in a superficial explanation, the motion equations are decoupled to obtain new solutions.  Notably, the distribution of matter in the bulk doesn't globally exhibit a negative cosmological constant. We also expect that the radial singularity is excluded and, finally, the parameter $a$ will influence the stability of the background metric. The investigation singularities will be done in the next subsections. The study of energy conditions and stability will be left to the next sections.

\subsection{Ricci and Kretschmann invariants}

A good way to study singularity is by evaluating the Ricci and Kretschmann invariants. They are respectively given by
\begin{equation} \label{RicciScalar1}
    R = -\frac{1}{l^2}\left(20 + \frac{2a^2z^2(\tilde{r} - 3M)}{\tilde{r}^5}\right).
\end{equation}
and 
\begin{align}\label{kreschamann}
    \mathcal{K} =&  \frac{1}{l^4}\left[40 + \frac{48M^2z^4}{\tilde{r}^6} + a^2\left(-\frac{24Mz^2}{\tilde{r}^5} + \frac{8z^2}{\tilde{r}^4} - \frac{144M^2z^4}{\tilde{r}^8} + \frac{32Mz^4}{\tilde{r}^7}\right) \right.\nonumber\\
    &  \left.+ a^4\left(\frac{132M^2z^4}{\tilde{r}^{10}} - \frac{64Mz^4}{\tilde{r}^9} + \frac{12z^4}{\tilde{r}^8}\right)\right]
\end{align}
It is important to note that, as expected, taking $a \to 0$ causes the central singularity at $r \to 0$ to reappear. Here we assume the general case where $a \neq 0$, so that we have the following asymptotic behaviors
\begin{equation} \label{eqRa}
    \lim_{\tilde{r} \to a} R= -\frac{1}{l^2}\left(20 + \frac{2z^2(a - 3M)}{a^3}\right),     
\end{equation}
\begin{equation}\label{eqKa}
    \lim_{\tilde{r} \to a} \mathcal{K} = \frac{1}{l^4}\left(40 + \frac{8z^2}{a^2} - \frac{24Mz^2}{a^3} + \frac{12z^4}{a^4} - \frac{32Mz^4}{a^5} + \frac{36M^2z^4}{a^6}\right),
\end{equation}
\begin{equation} \label{eqRi}
    \lim_{\tilde{r} \to \infty} R = -\frac{20}{l^2}, 
\end{equation}
\begin{equation} \label{eqKi}
    \lim_{\tilde{r} \to \infty} \mathcal{K} = \frac{40}{l^2}.
\end{equation}
From equations \eqref{eqRa}, \eqref{eqKa}, we can deduce that the spacetime $5D$ is regular at the origin $r=0$ (or $\tilde{r}=a$). On the other hand, from equations \eqref{eqRi} and \eqref{eqKi}, we observe that at infinity in the radial coordinate, the Ricci and Kretschmann scalars recover the values corresponding to an AdS space. Thus, our solution only behaves locally as AdS at the boundary of the asymptotic radial region.

In addition, we observe that $\mathcal{K}$ diverges as $z \to \infty$, which suggests the presence of a singularity at what is referred to as the AdS horizon. It is important to mention that in the case where the extra coordinate is compactified as $\bar{z} = \bar{R}z$, where $\bar{R}$ is the compactification radius, the new domain would be $\bar{z} \in [-\pi, \pi]$, thus suppressing the singularity at the AdS horizon.
In this context, the $\mathcal{Z}_2$ symmetry induces the presence of Dirac delta functions, located at the brane locations, in the equations of motion and consequently in the expression of the curvature scalar. This suggests that one interpretation could be that the presence of this Dirac delta distribution in the invariants does not imply a physical singularity like that produced by any diverging function, but represents a pathology associated with mirror symmetry and the confinement of matter. It is worth mentioning that singularities of functions, in general, are not volume integrable, unlike delta distributions, which are.

\subsection{Geodesic analysis}

In this section, we will carefully analyze the geodesic movement. We consider the geodesic equation with $\theta=\pi/2$ and study the $z$ and radial directions. For the $z$ coordinate we get
\begin{equation} \label{zgeodesic}
    \frac{d^2z}{d\lambda^2} - \frac{1}{z}U(\tilde{r})\left(\frac{dt}{d\lambda}\right)^2 + \frac{1}{z}U(\tilde{r})^{-1}\left(1 - \frac{a^2}{\tilde{r}^2}\right)^{-1}\left(\frac{d\tilde{r}}{d\lambda}\right)^2 + \frac{r^2}{z}\left(\frac{d\phi}{d\lambda}\right)^2 - \frac{1}{z}\left(\frac{dz}{d\lambda}\right)^2 = 0,
\end{equation}
where $\lambda$ is an affine parameter. Also, by the very definition of geodesics, $g_{MN}\dot{x}^M\dot{x}^N = \alpha$, where $\alpha = 0$ represents null geodesics and $\alpha = -1$ time-like geodesics. Developing we obtain
\begin{equation} \label{zgeodesic2}
    -\frac{1}{z}U(\tilde{r})\left(\frac{dt}{d\lambda}\right)^2 + \frac{1}{z}U(\tilde{r})^{-1}\left(1 - \frac{a^2}{\tilde{r}^2}\right)^{-1}\left(\frac{d\tilde{r}}{d\lambda}\right)^2 + \frac{1}{z}\tilde{r}^2\left(\frac{d\phi}{d\lambda}\right)^2 + \frac{1}{z}\left(\frac{dz}{d\lambda}\right)^2 = \frac{z\alpha}{l^2}.
\end{equation}
By subtracting Equations (\ref{zgeodesic}) and (\ref{zgeodesic2}), we have the equation that describes the movement in the $z$ direction,
\begin{equation}\label{zsimpsongeo}
    \frac{d}{d\lambda}\left(\frac{1}{z^2}\frac{dz}{d\lambda}\right) = \frac{\sigma}{zl^2},
\end{equation}
where $\sigma = 0$ for null geodesics and $\sigma = 1$ for time-like geodesics. 

For null and time-like geodesics, the solutions of this latter equation are given by
\begin{equation}\label{solutionz}
    z = -\frac{z_1l}{\lambda},  z = -z_1\csc(\lambda/l).
\end{equation}

In both cases, $z_1$ is a fixed constant such that $\displaystyle \lim_{\lambda \to 0^-} z \to \infty$. Thus, we can deduce that even with the radial regularization of Simpson-Visser, an object would be destroyed upon reaching the infinity of the radial coordinate.
 It is interesting to point out that this is consistent with the analysis of stability analysis in the next section. 

We can use the geodesic equation for $z$ to analyze the movement in the radial direction. As the metric does not depend on $t$ and $\phi$, the associated Killing vectors give the conservation of energy and angular momentum. Since $g_{00} = -l^2U(\tilde{r})/z^2$ and $g_{33} = \tilde{r}^2l^2/z^2$ (at the equator), then we get
\begin{equation}\label{killing}
    \frac{dt}{d\lambda} = \frac{Ez^2}{U(\tilde{r})l^2},\;  \frac{d\phi}{d\lambda} = \frac{Lz^2}{\tilde{r}^2l^2}.
\end{equation}
Thus, using $z = -\frac{z_1l}{\lambda}$ in Equation (\ref{zgeodesic}) and using equation \eqref{killing}, we have
\begin{equation}
    \left(\frac{d\tilde{r}}{d\lambda}\right)^2 + \frac{z^4}{l^4}\left[\left(\frac{l^2}{z_1^2} + \frac{L^2}{\tilde{r}^2}\right)U(\tilde{r}) - E^2\right]\left(1 - \frac{a^2}{\tilde{r}^2}\right) = 0.
\end{equation}
Now, it is convenient to return to the original variable $r$ given by $\tilde{r}^2 = r^2 + a^2$. This leaves the radial equation
\begin{equation}
     \left(\frac{dr}{d\lambda}\right)^2 + \frac{z^4}{l^4}\left[\left(\frac{l^2}{z_1^2} + \frac{L^2}{r^2 + a^2}\right)\left(1 - \frac{2M}{\sqrt{r^2 + a^2}}\right) - E^2\right] = 0
\end{equation}
By using the definitions 
\begin{equation}\label{parameters}
  \nu = -\frac{z_1^2}{\lambda},\;  \nu = - (z_1^2/l)\cot(\lambda/l)
\end{equation}
for null and time-like geodesics and define new coordinates and constants
\begin{equation}\label{coordinates}
 r = z_1r/l,\; \bar{E} = z_1E/l,\; \bar{L} = z_1^2L/l^2,\;\bar{M} = z_1M/l.
\end{equation}
and defining $\bar{a} = z_1a/l$, we get the radial equation
\begin{equation} \label{geodesicaSV1}
    \left(\frac{d\bar{r}}{d\nu}\right)^2 + \left(1 + \frac{\bar{L}^2}{\bar{r}^2 + \bar{a}^2}\right)\left(1 - \frac{2\bar{M}}{\sqrt{\bar{r}^2 + \bar{a}^2}}\right) = \bar{E}^2.
\end{equation}
This equation is equivalent to the radial equation of time-like geodesics with mass $\bar{M} = Mz_1/l$ and regularization parameter $\bar{a} = a z_1/l$ in a Simpson-Visser like geometry \cite{Simpson_2019}. According to \cite{Chamblin:1999by}, $\bar{M}$ is the ADM mass for an observer at $z = z_0 = l^2/z_1$. Once again, we can view it as the usual radial geodesic equation $\dot{\bar{r}}^2+V_{eff}=\bar{E}^2$. For $\bar{L}=0$, the singularity is excluded, in comparison with the case where the embedded $4D$ geometry corresponds to that of Schwarzschild.

We can see that the behavior near the singularity in $z$ holds. To conclude this, it's enough to calculate the Riemann scalar in a reference frame that propagates parallel to the geodesics. Let the tangent vector to these geodesics with $L = 0$ be given by
\begin{equation}
    u^M = \left(\frac{Ez^2}{\mathcal{U}(r)l^2},\frac{z^2}{l^2}\sqrt{E^2 - \frac{l^2}{z_1^2}\mathcal{U}(r)}, 0,0 ,  \frac{z}{l}\sqrt{\frac{z^2}{z_1^2} - 1} \right),
\end{equation}
where $\mathcal{U}(r) = 1 - \frac{2M}{\sqrt{r^2 + a^2}}$. The unit normal vector to the geodesic $n^M$, which is parallel propagated along it, is given by
\begin{equation}
    n^N = \left(-\frac{zz_1}{l^2\mathcal{U}(r)}\sqrt{E^2 - \frac{l^2}{z_1^2}\mathcal{U}(r)},-\frac{Ez_1z}{l^2},0,0,0\right).
\end{equation}
It is straightforward to check that $u^P\nabla_P n^Q = 0$. Thus, by calculating one of the curvature components in this reference frame, we obtain
\begin{equation} \label{InvarianteSV}
    R_{(u)(n)(u)(n)} = R_{MNPQ}u^Mn^Nu^Pn^Q = \frac{1}{l^2}\left[1 + \frac{z^4}{2z_1^2}\left(\frac{2M}{(r^2 + a^2)^{3/2}} - \frac{6Mr^2}{(r^2 + a^2)^{5/2}}\right)\right],
\end{equation}
which  diverges along the geodesic to $\lambda \to 0^-$. Hence, our case remains regular at the origin; however, there is still a divergence at the AdS horizon. However, the presence of a singularity at infinity in the radial coordinate could make sense from a physical perspective by considering the behavior of the geometry at the radial boundary. At this boundary, our solution behaves like an AdS space. On the other hand, it is well-known that the AdS radius acts as a sort of box, leading to instabilities near the infinity of the extra coordinate \cite{Gregory:1993vy,Chamblin:1999by}.

\section{\bf Energy Conditions and Stability}

In this section, we will study energy conditions and stability. For this, we need of the bulk components of the energy-momentum tensor \cite{Binetruy:1999ut}. Using the components of the Einstein tensor computed previously in Eq. (\ref{ET}), these components are
\begin{eqnarray}
   \rho_E &=& -\frac{1}{l^2\kappa_5^2}\left(6 + \frac{a^2z^2(\tilde{r} - 4M)}{\tilde{r}^5}\right); \;\;p_1 = \frac{1}{l^2\kappa_5^2}\left(6 - \frac{a^2z^2}{\tilde{r}^4}\right); \nonumber\\
   p_2 &=& \frac{1}{l^2\kappa_5^2}\left(6 +\frac{a^2z^2(\tilde{r} - M)}{\tilde{r}^5}\right); \;\;\; p_3 = \frac{1}{l^2\kappa_5^2}\left(6 + \frac{a^2z^2(\tilde{r} - 3M)}{\tilde{r}^5}\right). \label{pressures}
\end{eqnarray}
For $a = 0$, we have $\rho_E = - p_1= - p_2 = - p_3 = -6/(l^2\kappa_5^2) \equiv \Lambda_{5D}$, which is the negative cosmological constant term.

It is direct to check that the induced metric on the brane is given by the four-dimensional SV spacetime. The energy-momentum tensor located at the brane location $y=0$ in the original coordinates is given by $T^\mu_\nu=\delta(y) \text{diag}(\sigma,\sigma,\sigma,\sigma)$, where $\sigma$ is the brane tension. Given the absence of sources or matter fields on the brane, we can assert that the induced Simpson-Visser (SV) geometry on the brane should result from the influence of geometrical and matter deformations in the bulk. Therefore, it is not necessary to include additional matter fields at the location of the brane beyond its own tension. Using the junction conditions,  for the line element \eqref{elementodelineafinal1}, we can check that the brane tension is given by $\sigma=\dfrac{6}{\kappa^2 l}>0$.

\subsection{\bf Energy conditions}

First of all, we note that we will use the following definitions to analyze the energy conditions:
\begin{align}
\text{Null Energy Condition (NEC): }& \,  \rho + p_1 \geq 0 , \rho + p_2 \geq 0 , \rho + p_3 \geq 0 \\
\text{Weak Energy Condition (WEC): }& \, \rho \geq 0, \rho + p_1 \geq 0,  \rho + p_2 \geq 0,  \rho + p_3 \geq 0
\\
\text{Strong Energy Condition (SEC): }& \rho + p_1 \geq 0,  \rho + p_2 \geq 0,  \rho + p_3 \geq 0, \nonumber \\
& \rho+p_1+2p_2+p_3 \geq 0
\end{align}

With the expressions of the energy density and pressures obtained, we now examine if the energy conditions are globally satisfied in the bulk. 

\subsubsection{\bf Energy Density}

\begin{figure}[h]
    \centering
    \includegraphics[scale=.9]{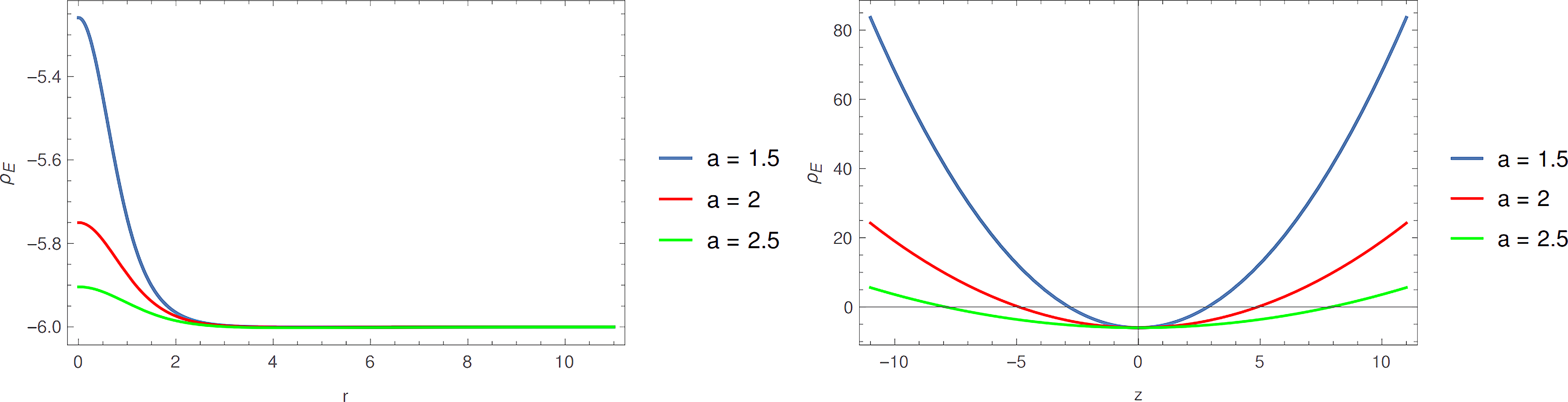}
    \caption{Energy density $\rho_E$ of the bulk for $l = \kappa_5 = M = z = 1$ (left) for the RBH case ($a < 2M$), OWWH case ($a = 2M$) and TWH case ($a > 2M$) and for $l = \kappa_5 = M = 1$ and $r = 0$ (right) for the RBH case ($a < 2M$), OWWH case ($a = 2M$) and TWH case ($a > 2M$).}
    \label{wec}
\end{figure}

In Fig.\ref{wec}, we present the plot of the energy density as a function of $r$ for a fixed point in the bulk. Analyzing the graph, it is evident that $\rho_E$ is globally negative for a fixed and finite $z$, thereby violating the Weak Energy Condition (WEC) in all three cases under consideration for this region. However, for a fixed $r$, the energy density can take positive values. In fact, for $r$ values close to $r = 0$, the energy density is positive for some values of $z$ such that $|z| > z_1$.

\subsubsection{\bf NEC}

Regarding the null energy condition (NEC), i.e., $\rho_E + p_i \geq 0$, we have the following expressions
\begin{eqnarray}
   \rho_E + p_1 &=&  -\frac{2a^2z^2}{l^2\kappa_5^2}\left(\frac{(r^2 + a^2)^{1/2} - 2M}{(r^2 + a^2)^{5/2}}\right), \label{necradial} \\
   \rho_E + p_2 &=& \frac{3a^2z^2}{l^2\kappa_5^2}\cdot\frac{M}{(r^2 + a^2)^{5/2}}, \\
  \rho_E + p_3 &=& \frac{a^2z^2}{l^2\kappa_5^2}\cdot\frac{M}{(r^2 +a^2)^{5/2}}. 
\end{eqnarray}

\begin{figure}[h]
    \centering
    \includegraphics[scale=.6]{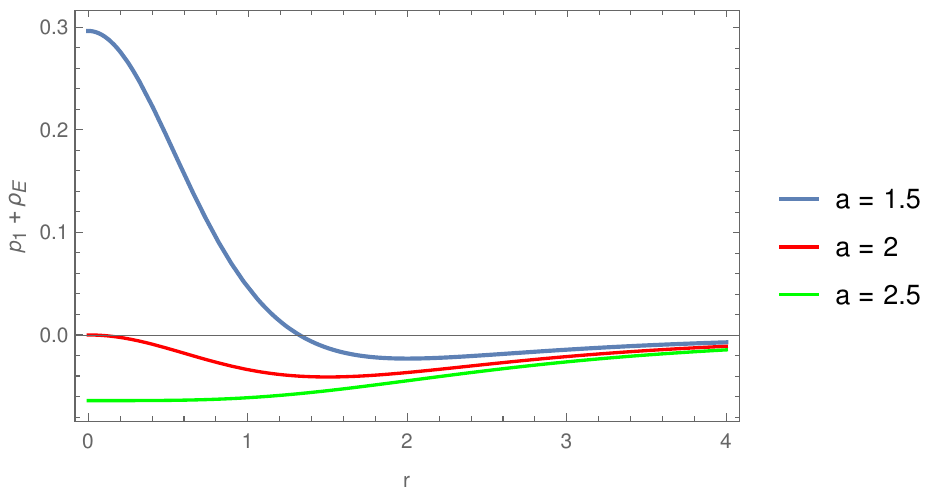}
    \caption{$\rho_E + p_1$ of the bulk for $l = \kappa_5 = M = z = 1$ for the RBH case ($a < 2M$), OWWH case ($a = 2M$) and TWH case ($a > 2M$).}
    \label{nec}
\end{figure}

From the above expressions, it is clear that the conditions $\rho_E + p_2 \geq 0$ and $\rho_E + p_3 \geq 0$ are satisfied for all values of $r$ and $z$. In Fig.~\ref{nec}, we plot $\rho_E + p_1$ as a function of $r$ for a fixed value of $z$. It is evident from this plot that the NEC is satisfied only within a specific region for the RBH case and at the asymptotic radial limit of $\rho_E + p_1$. Therefore, according to Equation \eqref{necradial}, the NEC is satisfied in this case only for $r \leq \sqrt{(2M)^2 - a^2}$, which corresponds to $r \leq r_h$ for the RBH case and in the limit as $r \to \infty$.

\subsubsection{\bf SEC}

From the statement of this subsection, we can observe that for the Strong Energy Condition (SEC) to be satisfied, both the Null Energy Condition (NEC) and the following additional condition must be met:
\begin{equation}\label{eqsec}
   \rho_E + \sum_i p_i = \frac{1}{l^2\kappa_5^2}\left(18 + \frac{a^2z^2}{\tilde{r}^5} (\tilde{r} - M)\right) \ge 0
\end{equation}

\begin{figure}[h]
    \centering
    \includegraphics[scale=.86]{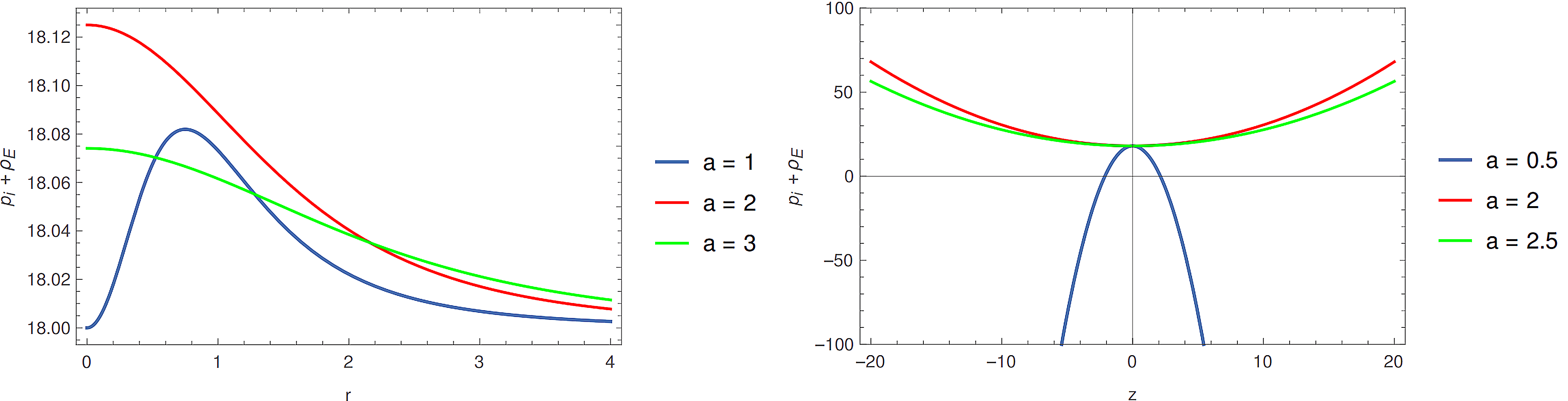}
    \caption{In the left panel, we have $\rho_E + \sum_ip_i$ of the bulk for $l = \kappa_5 = M = z = 1$ for the RBH case ($a < 2M$), OWWH case ($a = 2M$) and TWH case ($a > 2M$), while in the right panel we have the same plot of the expression for $l = \kappa_5 = M = 1$ and $r = 0$ or the RBH case ($a < 2M$), OWWH case ($a = 2M$) and TWH case ($a > 2M$)..}
    \label{sec}
\end{figure}

In the previous section, it was determined that the Null Energy Condition (NEC) is satisfied only for $\tilde{r} \leq 2M$ in the RBH case and in the limit as $r \to \infty$. On the Figure \ref{sec}, we observe that the condition \eqref{eqsec} can be satisfied for $a<2M$, corresponding to the RBH case. More rigorously, from Equations \eqref{necradial} and \eqref{eqsec}, the Strong Energy Condition (SEC) at $a<2M$ is satisfied at  for the cases where $\tilde{r} \to \infty$ and where:
\begin{equation}
  \tilde{r} \leq 2M  \,\, \land \,\, 18 + \frac{a^2z^2}{\tilde{r}^5}(\tilde{r} - M) \ge 0
\end{equation}

As mentioned in the introduction, observations of the shadows of M87* and the diameter of the image of Sgr A* have provided indications that the induced geometry on the brane could represent both ultra-compact objects and wormholes. Specifically, reference \cite{Banerjee:2022jog} argues that, although wormholes generally require exotic matter to remain stable, this is not necessarily the case in the presence of extra dimensions. They demonstrated that the wormhole solution they considered does not require exotic matter {\it on the brane}. Similarly, in our case, no additional exotic matter is needed on the brane beyond its own tension $\sigma$. However, since the energy conditions are not satisfied in several scenarios within the bulk, the inclusion of exotic matter in the bulk is necessary to sustain the induced geometry on the brane.

\subsection{Stability of the solution}

\subsubsection{\bf Discussion about wavelength perturbations and our geometry}

It is well known that black string solutions in asymptotically flat space are unstable to wavelength perturbations \cite{Gregory:1993vy}. As noted by \cite{Chamblin:1999by}, an AdS spacetime acts as a confining box, preventing the development of fluctuations with wavelengths much larger than $l$, thereby mitigating this type of instability. In this connection, if the curvature scalar $R = -20/l^2$ is very small, the AdS curvature becomes negligible, and the spacetime behaves like flat space. This implies that the black string solution will be unstable to perturbations with wavelengths on the order of the horizon radius $r_e = \frac{2Ml}{z_0}$, such that $\frac{r_e}{l} = \frac{2M}{z_0} \ll 1$. In other words, for very large values of $z$, such perturbations will fit within the AdS box, and instability could occur near the AdS horizon. However, for small values of $z_0$, potential instability would occur at wavelengths much larger than $l$, i.e., $\frac{r_e}{l} = \frac{2M}{z_0} \gg 1$, and thus the AdS curvature prevents instability. Therefore, the black string solution is stable far from the AdS horizon but potentially unstable near it. This issue was addressed when it was shown that RS black holes are unstable under linear perturbations \cite{Gregory:2000gf}.

In contrast to the previous paragraph, our regularized solution only exhibits AdS behavior at the asymptotic end of the radial coordinate. Thus, our spacetime is not nearly flat for large values of $l$ at finite radial coordinate values, as is the case in an AdS black string. Therefore, it is not accurate to claim that our solution is unstable near the AdS horizon and stable far from it. Such assertions would only be valid at the radial coordinate boundary where our geometry behaves like an AdS space. This makes it intriguing to study the stability under wavelength perturbations for finite radial coordinate values. However, this investigation extends beyond the scope of our current research and could be explored in future work.

\subsubsection{\bf Speed of sound criteria}

The fact that the criterion described in the previous section appears to be incompatible with our geometry prompts us to consider alternative criteria. In this context, it is well known that the stability criterion and causality associated with the speed of sound have been applied in the literature to solutions featuring matter in the energy-momentum tensor \cite{Capozziello:2022zoz,Das:2020lct}. Specifically, the speed of sound in a direction $i$ is defined as:
\begin{equation}\label{vsounddef}
    (v^{i}_s)^2=\frac{dp_i}{d\rho},
\end{equation}
See, for example, Appendix II-A of Reference \cite{Armendariz-Picon:2005oog}, where the last formula is derived from perturbations of the Lagrangian dependent on the squared gradient of a scalar field. Also, refer to Chapter 1.7 of Reference \cite{Vietri+2008}, where the previous formula is derived by applying perturbations to the density and pressure in the mass conservation equation.

The stability condition states that $0 \leq (v^{i}_s)^2 \leq 1$ \cite{Capozziello:2022zoz,Das:2020lct}. Reference \cite{Ellis:2007ic} indicates that the situation where $(v^{i}_s)^2 > 1$ implies that the speed of sound cones lie outside the light cones in all directions at all events. Consequently, fluid waves can propagate at speeds up to and including this superluminal speed of sound. This reference also notes that such a situation is far from what is typically observed in ordinary matter.

First of all, we can observe that at infinity, the equation of state is recovered as $p_i \sim -\Lambda \sim \rho$. This is because, at infinity, the space exhibits an AdS (Anti-de Sitter) geometry. Consequently, in the parameter space, we can deduce that $\frac{d<p>}{d\rho} = -1$. As noted in reference \cite{Giambo:2023zmy}, it is important to emphasize that the sound speed represents the velocity of perturbations in a given fluid. In the case of negative equation of state parameters, it is possible for the squared sound speed to become negative, leading to instabilities.

Thus, in the context of braneworld models with the presence of matter fields in the bulk, we propose to study stability based on the causality criterion of the speed of sound. To do this, we will use the assumption that all pressures propagate uniformly in all spatial directions of the bulk. In this way, We will now investigate the stability of the obtained braneworld black bounce solutions by analyzing the sound velocities of the fluid along radial and extra-dimensional directions. From equations (\ref{vsounddef}) and  (\ref{pressures}) we get
\begin{eqnarray}\label{vsound}
    (v^{r}_s)^2&=&\frac{d <p>}{d\rho}=\frac{<p>'}{\rho'}=\frac{25 M-8 \sqrt{a^2+r^2}}{16 \left(\sqrt{a^2+r^2}-5 M\right)}\\
    (v^{z}_s)^2&=&\frac{<p>'}{\rho'}=\frac{5 M-2 \sqrt{a^2+r^2}}{4 \left(\sqrt{a^2+r^2}-4 M\right)}
\end{eqnarray}
in the coordinates space, using the prime to denote the derivative concerning coordinate $r$ (coordinate $z$). We have used the fact that the average pressure $<p>$ is given by
\begin{equation}\label{averpres}
<p>=\frac{1}{4}[p_1+2p_2+p_3],
\end{equation}
since some lateral pressures are equal. Note that $v^{i}_s$ is independent of the extra-dimensional coordinate, $z$. This should be expected since the motion along the extra dimension is also independent of $z$. 

\begin{figure}[t]
    \centering
    \includegraphics[scale=.65]{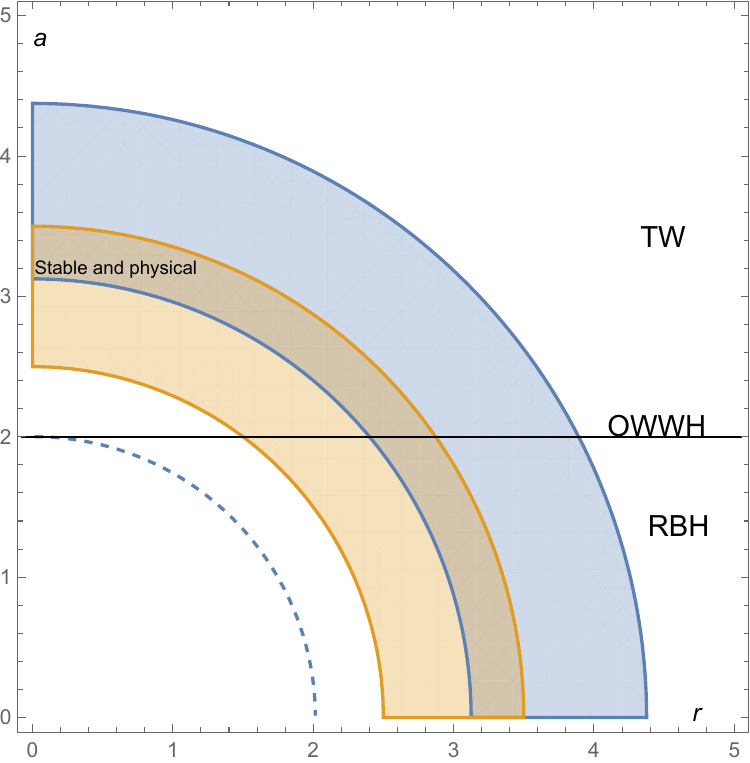}
    \caption{Parameter space $(r,a)$ depicting regions of stable and physical solutions, considering $M = 1$ and $a\neq 0$, for the RBH case ($a < 2M$), OWWH case ($a = 2M$), and TWH case ($a > 2M$). The dashed line denotes the RBH horizon. The superposition region between blue (along $r$ direction) and orange (along $z$ direction) circular sectors indicates where the sound velocities obey $0\leq (v_s^{i})^2<1$.}
    \label{stab}
\end{figure}

As was mentioned, to ensure stability in the solution, we must impose $0\leq (v^{i}_s)^2\leq 1$, or
\begin{equation}
    0\leq\frac{25 M-8 \sqrt{a^2+r^2}}{16 \left(\sqrt{a^2+r^2}-5 M\right)}\leq 1; \text{ } 0 \leq \frac{5 M-2 \sqrt{a^2+r^2}}{4 \left(\sqrt{a^2+r^2}-4 M\right)}\leq1.
\end{equation}
With this, we can study the possibility of a stable solution.  For traversable wormholes ($a > 2M$) with throat radii in the interval $\frac{25}{8}M\leq a \leq \frac{7}{2} M$, they are stable and physical, at least near and at the throat ($r\approx 0$). In contrast, OWWHs exhibit instability in these regions. Conversely, for RBHs, where $a<2M$, stability and physical validity dominate in a narrow region beyond and outside the horizon. Additionally, the entirety of their inner regions fails to meet those criteria. We plot the region of stability in Fig. \ref{stab}. 
Since we have most of the regions with instability, the conclusion is that solutions persist to be unstable. However, it is worth noting that we have some stable regions, which is a very different result compared to the vacuum AdS case.  

It is important to note that, as might be expected, instabilities also occur in the coordinate space. This can be easily observed by taking the limits $\displaystyle \lim_{r \to \infty} (v^{r}_s)^2 = \lim_{r \to \infty} (v^{z}_s)^2 = -\frac{1}{2}$. This is expected because, at the boundary, the geometry behaves like an AdS (Anti-de Sitter) space.

\section{Conclusion and Summarize}

In this work, we have studied the Simpson-Visser regularization of the Randall-Sundrum Schwarzschild black hole. To accomplish this, we linearly deform the bulk geometry, represented by the $5D$ Einstein tensor, and the bulk matter distribution, represented by the $5D$ energy-momentum tensor, with respect to a coupling constant $a^2$. The seed solution corresponds to the $5D$ AdS spacetime given by the black string of Hawking et.al. \cite{Chamblin:1999by}. 

Moreover, both the geometrical and matter deformations, represented by equations \eqref{EinsteinDeformado} and \eqref{MateriaDeformado} respectively, result in the induced geometry on the brane transitioning from a singular Schwarzschild spacetime to a regularized SV spacetime. Given the absence of sources or matter fields on the brane, we can assert that the induced SV geometry on the brane arises from the influence of geometrical and matter deformations in the bulk. In this context, the $5D$ spacetime is regular at the origin $r=0$ (or $\tilde{r}=a$), unlike the AdS case, where singularity occurs at the origin. However, the singularity at the AdS horizon persists. This latter singularity could be suppressed in a scenario where the extra dimension is compactified.
Despite being asymptotically AdS, the above deformation induces a transition from a vacuum distribution to an anisotropic distribution.

Our study of the energy conditions reveals that the Strong Energy Condition (SEC) may or may not be satisfied depending on the parameter values. From the perspective of the brane itself, our results are consistent with observations of the shadows of M87* and the diameter of the image of Sgr A*. These observations suggest that the induced geometry on the brane could represent both ultra-compact objects and wormholes. Specifically, reference \cite{Banerjee:2022jog} argues that while wormholes generally require exotic matter to remain stable, this is not necessary in the presence of extra dimensions because exotic matter is not required {\it on the brane}. Similarly, in our case, no additional exotic matter is needed on the brane beyond its own tension $\sigma$. However, since the energy conditions are not satisfied in several scenarios within the bulk, the inclusion of exotic matter in the bulk is necessary to sustain the induced geometry on the brane.

On the other hand, since the spacetime is only AdS at the asymptote of the radial coordinate, we can conclude that the solution is unstable near the AdS horizon but stable far from it when considering wavelength perturbations criterion. This observation prompts us to investigate whether the solution remains stable at finite values of the radial coordinate, where the spacetime deviates from AdS. Consequently, we need to explore alternative criteria for assessing the stability of the geometry at these finite radial distances. In the context of braneworld models with matter fields in the bulk, we propose evaluating stability based on the causality criterion of the speed of sound. We have also analyzed the stability and physical validity of the braneworld black-bounce solutions by examining the sound speed along the $r$ and $z$ directions. Our analysis identified specific regions where stability and physicality are assured for both wormhole and regular black hole solutions, depending on the bounce parameter $a$, as illustrated in Fig. \ref{stab}. It is important to note that for braneworld models where $a = 0$, no stable regions are present. This aligns with the wavelength perturbation criterion because, in the limit $a \to 0$, the spacetime behaves as AdS. These instabilities are also observed in the limit $r \to \infty$, where the spacetime again behaves as AdS.

Finally, we mention that in the case of the coordinate $z$, we cannot do a Simpson-Visser-like regularization  $z \to \sqrt{z^2 + b^2} $, as we would continue to have a divergence in $z \to \infty $. An alternative would be to consider an exchange of the type $z \to f_b (z)$ so that $\displaystyle \lim_{z \to \infty} f_b(z)$ is a constant and $\displaystyle \lim_{b \to 0} f_b(z) = z$. However, nothing guarantees that such a function exists, and if it does, imposing an analytical form for $f_b(z)$ does not seem to be trivial. It is also necessary to test whether the regularized black string solution is unstable for finite values of the radial coordinate and at the AdS horizon, where the space-time is no longer flat, as in the non-regularized case. These aspects will be explored in future work.

\section*{Acknowledgements}
 Milko Estrada is funded by ANID , FONDECYT de Iniciaci\'on en Investigación 2023, Folio 11230247. Tiago M. Crispim, Geová Alencar, and Celio R. Muniz would like to thank Conselho Nacional de Desenvolvimento Científico e Tecnológico (CNPq) and Fundação Cearense de Apoio ao Desenvolvimento Científico e Tecnológico (FUNCAP) for the financial support.

\bibliography{ref.bib}

\end{document}